\documentclass[aps,prx,superscriptaddress,notitlepage,twocolumn]{revtex4-1}

\usepackage{graphicx}
\usepackage{amsmath,amssymb,amsfonts}
\usepackage{braket}
\usepackage{comment}
\usepackage{xcolor}
\usepackage{url}
\usepackage[hidelinks]{hyperref}
\usepackage{natbib}
\usepackage{dcolumn}
\usepackage{bm}

\newcommand{\UIUC}{Department of Physics, IQUIST, University of Illinois, IL 61801, USA.}
\newcommand{\JQI}{Present address: Joint Quantum Institute and Joint Center for Quantum Information and Computer Science, University of Maryland and NIST, College Park, Maryland 20742, USA}
\newcommand{\JILA}{JILA, National Institute of Standards and Technology and Department of Physics, University of Colorado, Boulder, CO, 80309, USA}
\newcommand{\CTQM}{Center for Theory of Quantum Matter, University of Colorado, Boulder, CO, 80309, USA}
\newcommand{\MIT}{Present address: Massachusetts Institute of Technology, Department of Physics, Cambridge, MA 02139}
\newcommand{\CUB}{Department of Physics, University of Colorado, Boulder, CO 80309, USA}
\newcommand{\contrib}{\thanks{Authors W.M.~and S.R.M.~contributed equally to this work.}}

\begin{document}

\title{Disorder-controlled relaxation in a 3D Hubbard model quantum simulator}

\author{W.~Morong} \contrib
\affiliation{\UIUC}
\affiliation{\JQI}
\author{S.~R.~Muleady} \contrib
\author{I.~Kimchi}
\affiliation{\JILA}
\affiliation{\CTQM}
\author{W.~Xu}
\affiliation{\UIUC}
\affiliation{\MIT}
\author{R.~M.~Nandkishore}
\affiliation{\CTQM}
\affiliation{\CUB}
\author{A.~M.~Rey}
\affiliation{\JILA}
\affiliation{\CTQM}
\author{B.~DeMarco}
\affiliation{\UIUC}

\begin{abstract}
Understanding the collective behavior of strongly correlated electrons in materials remains a central problem in many-particle quantum physics. A minimal description of these systems is provided by the disordered Fermi-Hubbard model (DFHM), which incorporates the interplay of motion in a disordered lattice with local inter-particle interactions. Despite its minimal elements, many dynamical properties of the DFHM are not well understood, owing to the complexity of systems combining out-of-equilibrium behavior, interactions, and disorder in higher spatial dimensions. Here, we study the relaxation dynamics of doubly occupied lattice sites in the three-dimensional (3D) DFHM using interaction-quench measurements on a quantum simulator composed of fermionic atoms confined in an optical lattice. In addition to observing the widely studied effect of disorder inhibiting relaxation, we find that the cooperation between strong interactions and disorder also leads to the emergence of a dynamical regime characterized by \textit{disorder-enhanced} relaxation. To support these results, we develop an approximate numerical method and a phenomenological model that each capture the essential physics of the decay dynamics. Our results provide a theoretical framework for a previously inaccessible regime of the DFHM and demonstrate the ability of quantum simulators to enable understanding of complex many-body systems through minimal models.

\end{abstract}

\maketitle

\section{Introduction}

\begin{figure*}[!thb]
\centering
\includegraphics[width=0.8\textwidth]{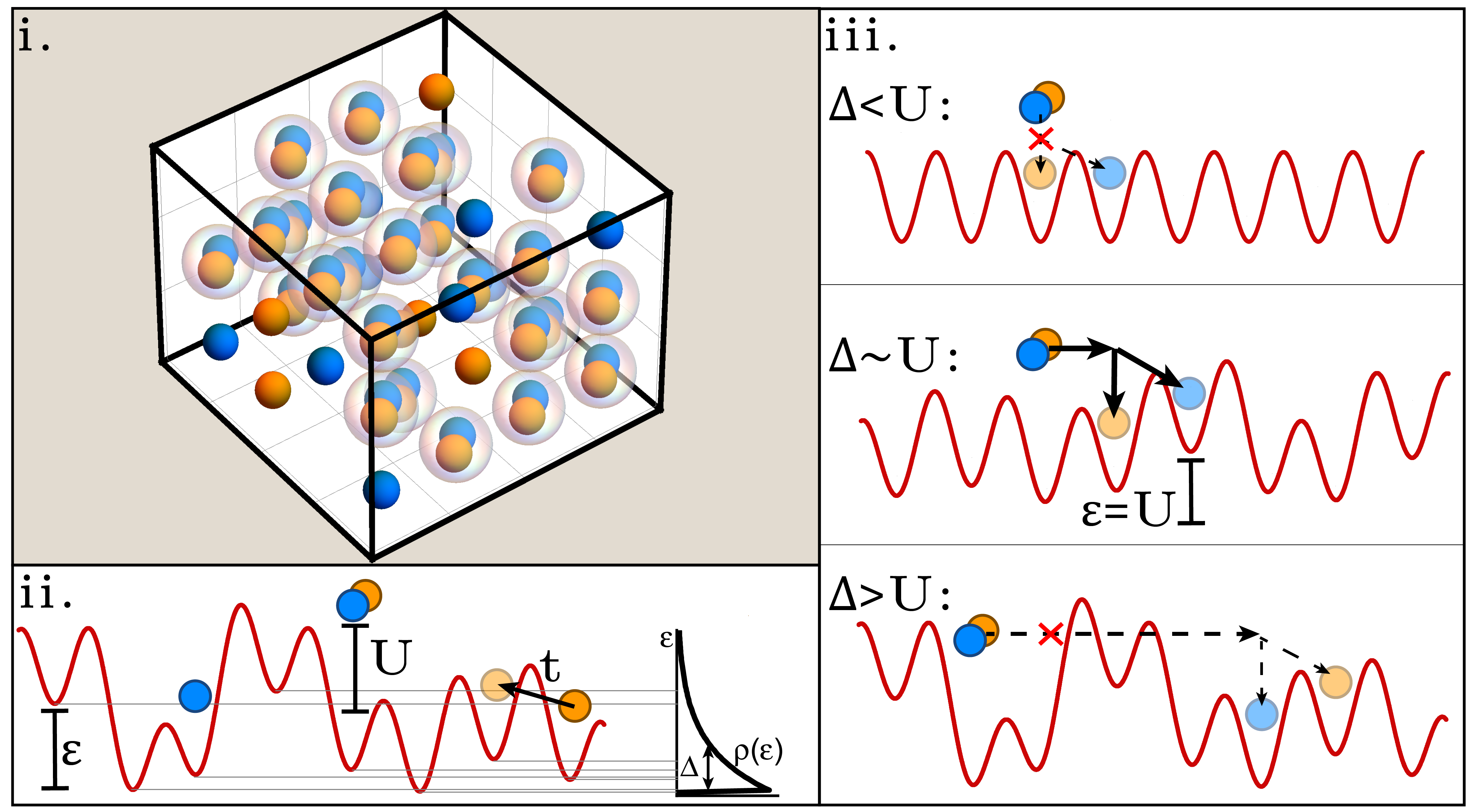}
\caption{Experimental setup. i: A nonequilibrium doublon population for atoms in two spin states is prepared in a 3D cubic lattice using an interaction quench. ii: The atoms are described by the disordered Fermi-Hubbard model (Eq. \ref{eq:DFHM}), which involves tunneling with amplitude $t$, doublon formation with interaction energy $U$, and a local disordered energy offset $\epsilon$ with an exponential distribution $\rho\left(\epsilon\right)$ characterized by disorder strength $\Delta$. iii: While the full dynamics of the system involve the interplay of doublons and single atoms, the case of an isolated doublon is useful for intuition about the dynamical regimes. For weak disorder (compared to $U$), the decay of doublons into pairs of single atoms is suppressed by the gap between the doublon and single-particle energies. Moderate disorder can enhance the decay rate by creating resonant pairs of sites with an energy difference comparable to $U$. However, sufficiently strong disorder can hinder decay by suppressing diffusion to these resonant sites.}
\label{fig:Fig1}
\end{figure*}

Strong disorder and interactions are known to give rise to the celebrated Anderson and Mott metal--insulator transitions. In an Anderson insulator, a random spatial potential localizes non-interacting particles through destructive interference \cite{Anderson1958,Lee1985}, while for a unit-filled Mott insulator, strong repulsive interactions create an energy gap that prevents particle motion \cite{Mott1949}. The combined presence of disorder and interactions in many physical systems poses the open challenge of understanding the interplay between these vastly different localization mechanisms \cite{Belitz1994, byczuk2010anderson}. The development of highly tunable and isolated quantum simulators, such as ultracold atoms trapped in optical lattices, has created new opportunities to experimentally study this long-standing problem using a minimal model combining both elements: the disordered Fermi-Hubbard model (DFHM) \cite{Sanchez-Palencia2010}.

A potential result of combined disorder and interactions in isolated systems is many-body localization (MBL), in which relaxation to thermal equilibrium is prevented by sufficiently strong disorder \cite{Gornyi2005,Basko2006,Nandkishore2014,Abanin2019}. Despite a concerted theoretical effort in recent years and several experimental results for one-dimensional chains \cite{Schreiber2015,Smith2016,Roushan2017,Luschen2017a}, many questions still remain regarding the nature of this phenomenon. This shortcoming is especially true for systems with more than one spatial dimension, although initial experimental studies with atoms in two and three-dimensional optical lattices have also observed slow dynamics consistent with MBL \cite{Kondov2013a,Choi2016, Bordia2017}.

Another mechanism that can suppress relaxation emerges in the strongly interacting regime of the DFHM: the formation of quasi-bound doubly occupied sites (i.e., doublons), which slowly decay in a clean lattice through high-order processes that generate many low-energy excitations \cite{Strohmaier2010,Sensarma2010}. The effect of disorder on doublon relaxation is largely unexplored. Reconciling the interplay of slow dynamics caused by doublon binding and disorder-induced localization is critical to obtaining a more complete understanding of thermalization in the DFHM, including the possibility of MBL. However, advancing this frontier demands exploring the highly non-trivial regime characterized by comparable disorder and interaction energies.

Here, we investigate how strong interactions and disorder compete and cooperate to affect the far-from-equilibrium dynamics of the three-dimensional DFHM. Using a quantum simulator of fermionic atoms in an optical lattice, we perform measurements of doublon relaxation following an interaction quench and use the resulting decay times to characterize the system behavior. Exploring the parameter regime from interaction-dominated to disorder-dominated behavior, we are able to classify the dynamics in terms of two distinct regimes: disorder-suppressed relaxation at strong disorder, and disorder-enhanced relaxation at weaker disorder. The latter effect has not previously been observed in a quantum simulator platform and may be related to disorder-driven insulator-metal transitions observed in certain correlated materials \cite{Lahoud2014, Wang2018}. We compare our results to beyond-mean-field numerical simulations of the quantum dynamics, which capture many features of our observed results and suggest the creation of resonances in the lattice by the disorder as the physical origin of this disorder-enhanced regime. We further supplement this picture by developing a simple phenomenological model that incorporates both disorder-enhanced and disorder-suppressed mechanisms for doublon decay.

\section{Measuring and simulating relaxation}

We employ an optical lattice experimental platform (Fig.~\ref{fig:Fig1}) described in previous work \cite{Kondov2013a}. Two spin states (denoted $\left | \uparrow \right\rangle$ and $\left | \downarrow \right\rangle$) of fermionic $^{40}$K atoms are trapped in a cubic lattice superimposed with 532~nm optical speckle disorder. This system realizes the DFHM with confinement (Fig.~\ref{fig:Fig1}):
\begin{equation} \begin{split}H=\sum_{\langle i, j \rangle,\sigma} \left(-t_{ij} \hat{c}^\dagger _{j \sigma}c_{i \sigma} +h.c.\right)
+\sum_i U_i n_{i \downarrow} n_{i \uparrow} \\
+\sum_{i, \sigma}\left( \epsilon_{i} + \frac{1}{2} m \omega^2 r_i^2 \right) n_{i \sigma}.
\label{eq:DFHM}
\end{split}
\end{equation}
\noindent
Here, $\sigma$ indexes the two spin states, $t_{ij}$ is the tunneling energy between sites $i$ and $j$  (restricted to nearest-neighbors, indicated by $\langle i, j \rangle$), $U_i$ is the on-site interaction energy, $\epsilon_i$ is the local disordered energy offset, $\omega$ is the harmonic confinement, $m$ is the atomic mass, and $r_i$ is the  distance of site $i$ from the trap center. The single-particle bandwidth is $12t$ in the absence of disorder. The applied speckle potential creates disorder in the $t$, $U$, and $\epsilon$ terms, leading to distributions of these Hubbard parameters with widths that depend on the optical power \cite{Zhou2010, White2009b}. We characterize the disorder strength by $\Delta$, which is approximately equal to the standard deviation of the $\epsilon$ distribution (Fig. 1(ii)). The influence of spatial correlations in the disorder potential is weak, since the correlation length of the speckle field is smaller than two lattice spacings along every lattice direction. Because of the harmonic confinement, all measurements are averaged over a density profile that varies from an estimated occupancy $\left \langle n \right \rangle=0.5$ at the center of the trap to zero at the edges of the system. 

To probe far-from-equilibrium doublon dynamics, we measure the population of atoms in doubly occupied sites following a quench in which the interactions are reversed from attractive to repulsive using a Feshbach resonance. Before the quench, the gas is in equilibrium with an energetically favorable doublon population. Afterwards, the doublons become excitations that can decay by breaking apart into a pair of single atoms (``singles"). The atomic doublon population is allowed to evolve in a disordered lattice by turning on the optical speckle field following the interaction quench.  After a variable time, the doublon population is measured by mapping each doublon to a tightly bound Feshbach molecule and selectively transferring the $\left | \downarrow \right \rangle$ atom in each molecule to an ancillary spin state using an rf sweep (see Appendix \ref{appendix:experiment}). Alternatively, we can selectively transfer and image only atoms from singles, which allows us to separate doublon decay from overall number loss.

\begin{figure*}
\centering
\includegraphics[width=0.8\textwidth]{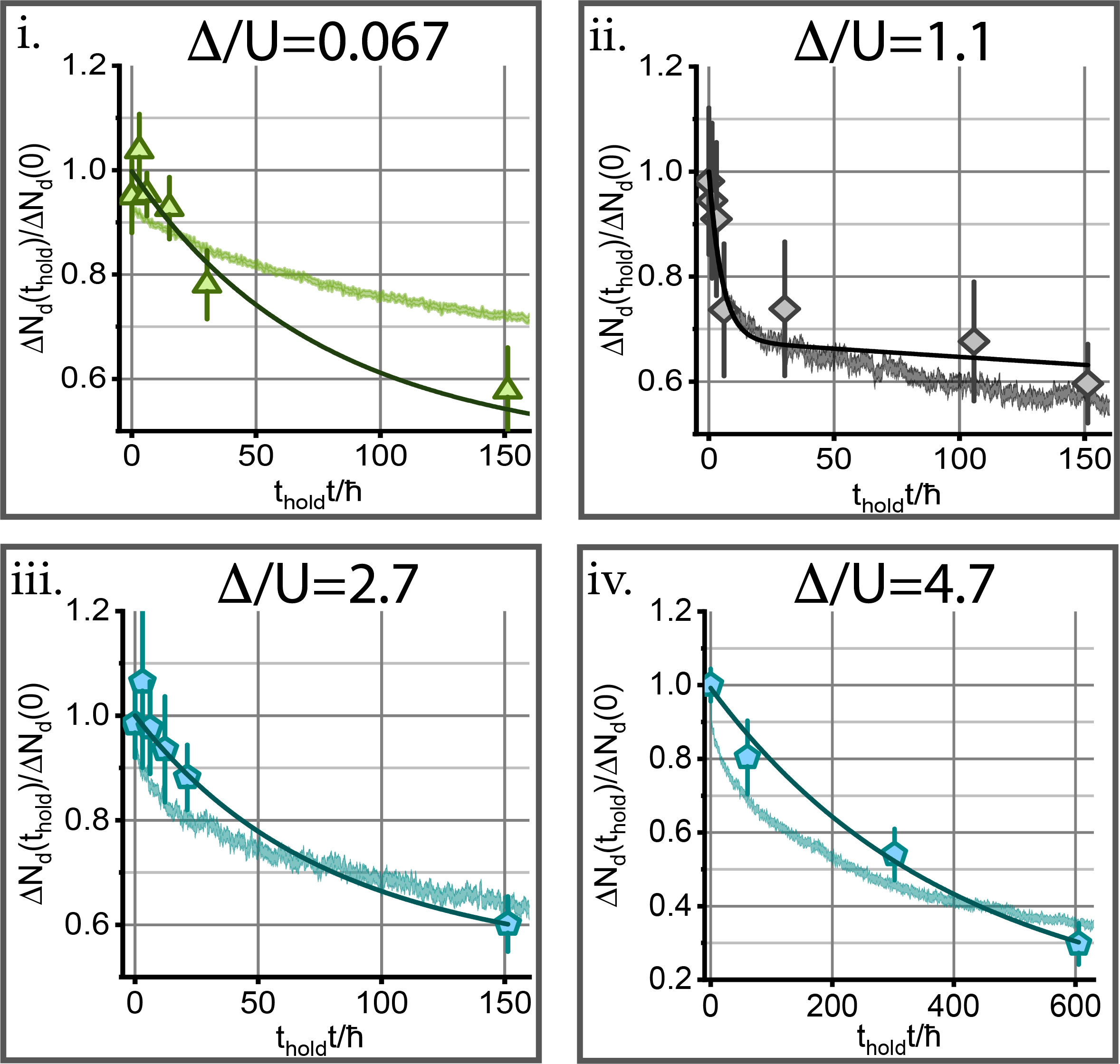}
\caption{Examples of experimental and numerical doublon relaxation data. Sample traces of the doublon population vs. time for different disorder strengths are shown for experimental measurements (points); the error bars give the statistical standard error of the mean for averaging over multiple measurements. The fits used to extract $\tau$ are shown using solid lines. The finite offset value at  long times  reflects the equilibrium doublon population; see Appendix \ref{appendix:experiment} and Fig. \ref{fig:FigS5a}. The corresponding GDTWA simulation (see Appendix \ref{appendix:gdtwa}) is displayed as a shaded region that shows the standard error of the mean from disorder and trajectory averaging. The horizontal axis for panel iv (the strongest disorder) is compressed to bring the slow decay into view, and all data are taken at $U/12t=1.8$ (corresponding to a $12$~$E_R$ lattice depth in the experiment).}
\label{fig:Fig2}
\end{figure*}

In all regimes, the doublon population decreases following the quench with a rate sensitive to the disorder strength. Typical results at different disorder strengths are shown in Fig. \ref{fig:Fig2}. To quantify the decay, we fit the data to a model that describes exponential doublon decay with a time constant $\tau$  and that includes overall particle number loss (see Appendix \ref{appendix:experiment}). While we find  that this fit provides a reasonable characterization of the decay timescale, the functional form of the relaxation is unknown beyond the clean limit. We therefore turn towards numerics to provide an interpretation of the timescale with disorder present.

The large scale and dimensionality of our system, as well as the far-from-equilibrium nature of the dynamics, preclude exact numerical studies and commonly employed approximate techniques, such as DMRG or 
diagonalization methods. Furthermore, the fermionic sign problem forbids use of quantum Monte Carlo methods. We therefore develop a numerical method---a generalized discrete truncated Wigner approximation (GDTWA) \cite{Zhu2019,Schachenmayer2015,Acevedo2017}---to simulate the relaxation process. The GDTWA approach invokes a factorization of the density matrix $\hat{\rho}$ of the many-body system over individual lattice sites $i$, $\hat{\rho} = \bigotimes_i \hat{\rho}_i$. Furthermore, random sampling of initial conditions from a discrete semiclassical phase space accounts for quantum noise.

More specifically, we can fully describe the reduced state $\hat{\rho}_i$ at lattice site $i$ (with a four-dimensional local Hilbert space spanned by basis states $\{\ket{\uparrow\downarrow},\ket{\uparrow},\ket{\downarrow},\ket{0}\}$) by a 16-dimensional vector $\vec{\lambda}_i$, so that $\hat{\rho}_i = \sum_{\alpha} \lambda_i^{\alpha}\hat{\Lambda}_i^{\alpha}$ for a complete basis of local observables $\{\hat{\Lambda}_i^{\alpha}\}$. Inserting the product state ansatz into the von Neumann equation $\partial_{\mathfrak{t}}\hat{\rho} = -i\hbar[\hat{H},\hat{\rho}]$ results in a set nonlinear differential equations describing the evolution of each $\vec{\lambda}_i$, and which may be numerically integrated to obtain the mean field dynamics. While this ansatz retains full information regarding the strong on-site Hubbard correlations, such a solution describes a product state at all instances in time and therefore neglects the important cross-site quantum correlations responsible for doublon decay. In fact, for an initial product state of definite particle number and spin, the mean field solution results in no dynamics for this system.

To capture the buildup of  quantum correlations between sites, we instead describe the initial value of each $\vec{\lambda}_i$ as a probability distribution over a discrete phase space that samples over all allowed  values for each observable in our local basis (see Appendix \ref{appendix:gdtwa}). Averaging independently evolved sets of randomly sampled initial conditions from this distribution yields a highly nontrivial solution for the dynamics (owing to the nonlinear nature of the dynamical equations) and describes the evolution to a correlated state exhibiting entanglement \cite{Zhu2019,Acevedo2017}. This approximation, while only rigorously valid at short times,  has demonstrated the ability to provide accurate results in generic spin models  at longer times and  properly  capture quantum thermalization of local observables \cite{Zhu2019,Lepoutre2019,Patscheider2020}. Here, for the first time, we adapt the GDTWA  to model DFHM dynamics and we provide benchmarks for its applicability  in Appendix \ref{appendix:gdtwa}.

The GDTWA results, shown as traces in Fig. \ref{fig:Fig2}, approximately capture the observed changes in doublon population decay with applied disorder. The GDTWA dynamics also allow us to determine, within the assumptions of the technique, the extent to which $\tau$ accurately characterizes the relaxation timescale. As illustrated in Fig. \ref{fig:Fig2},  we observe  qualitative differences between the measurements and simulations, which tend to exhibit a faster initial decay (over timescales $\hbar/t$) before transitioning to a slower decay on longer timescales that is well fit by an exponential decay. Nonetheless, by constraining the initial doublon fraction in the fitted exponential decay to the measured value, we are able to obtain a reasonable exponential fit of the entire dynamical curve that is used to extract a consistent relaxation timescale. The resulting effective timescale incorporates both the initial non-exponential features and the subsequent exponential decay (see Appendix \ref{appendix:gdtwa}, Fig. \ref{fig:FigS8}). 

To account for the spatially varying trap density and variations in the initial preparation, which affects the equilibrium doublon density, we choose initial singles densities for the simulations that yield best fits to the experimental data. These best-fit singles densities in the simulations are generally consistent with the experimentally measured single density at the trap center, except at small disorders where GDTWA does not fully capture the decay processes associated with the clean system and thus results in a much larger equilibrium doublon density for all initial singles densities, as evidenced in Fig.~2(i) (see also Appendix \ref{appendix:gdtwa}). However, we have also confirmed that the relaxation timescale is not strongly influenced by the average density nor the spatial  profile of the gas in the regime of interest (see Appendix \ref{appendix:gdtwa}, Fig. \ref{fig:FigS7}). These observations  support  characterizing the doublon relaxation dynamics by a single parameter $\tau$: $\tau$ primarily depends on the interplay of disorder with interactions and tunneling and is insensitive to parameters that may fluctuate or have some uncertainty in the experiment.

\section{Analysis of dynamical regimes}

\begin{figure*}
\centering
\includegraphics[width=0.8\textwidth]{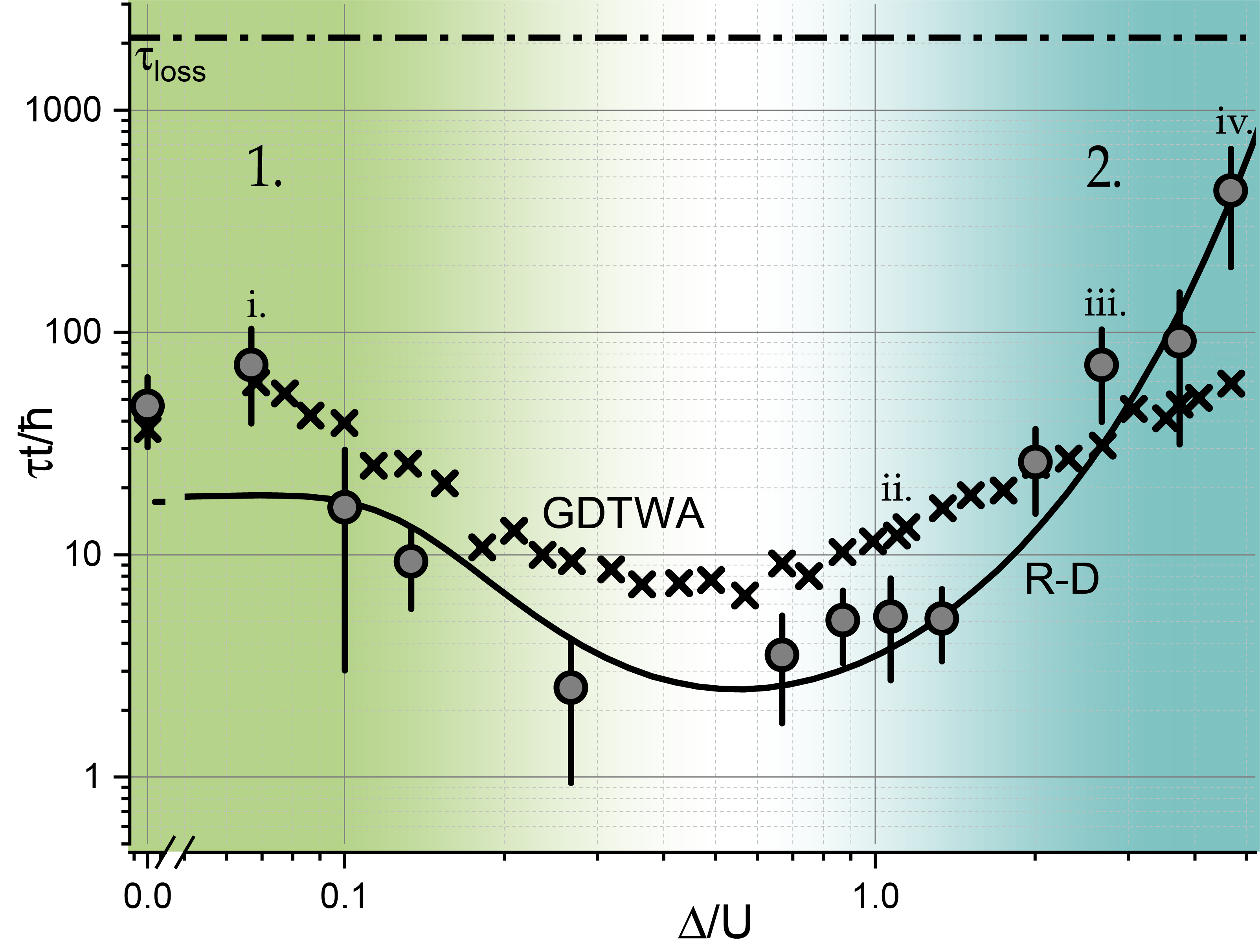}
\caption{Dynamical regimes of doublon relaxation. The doublon lifetime $\tau$ shows a strong  non-monotonic dependence on disorder, first decreasing (green region, labeled 1) and then increasing (blue region, labeled 2) with larger disorder strength. The measurements (points) are compared to numerical GDTWA simulations (crosses). The error bars for the experimental data provide the standard error of the fit used to determine $\tau$, while the error bars for the GDTWA numerical data are smaller than the symbols. The solid line represents a fit of the measurements to the analytical reaction--diffusion model with a diffusion constant that decreases monotonically with disorder. The lifetime for atom number loss in the experiment $\tau_{loss}$ (dash--dotted line) is independent of disorder and well separated from the doublon dynamics timescale. The labels i.--iv. indicate the values of $\Delta/U$ corresponding to panels i.--iv. of Fig. \ref{fig:Fig2}.
}
\label{fig:Fig3}
\end{figure*}

The dependence of $\tau$ on disorder is shown in Fig. \ref{fig:Fig3}; the qualitative features are shared by experiment and numerics. Strikingly, we find that applying disorder first causes the relaxation time $\tau$ to rapidly decrease. For the clean system, $\tau$ is substantially longer than the single-particle tunneling time $\hbar/t$, which is consistent with previous studies that identified doublons as repulsively bound pairs \cite{Winkler2006, Strohmaier2010, Sensarma2010, Shastry2011,Covey2016}. However, disorder causes $\tau$ to decrease to a minimum value comparable to the tunneling time $\hbar/t$ at a disorder value near $\Delta \sim U/2$. While the complementary phenomenon of interaction-driven delocalization has been observed in similar systems \cite{Deissler2010,Kondov2013a,Schreiber2015}, to our knowledge this is the first observation of a disorder-driven increase in relaxation in a quantum simulator with a high degree of isolation and tunability.

As $\Delta$ is increased beyond $U$, $\tau$ eventually increases, growing by over two orders of magnitude at the strongest disorder we can apply in the experiment. The separation into two dynamical regimes (distinguished by the slope of $\tau$ with $\Delta$) combined with the crossover at $\Delta \sim U$ suggest that the dynamics are controlled by competing mechanisms arising from interactions and disorder.

We can understand these dynamical regimes using a minimal model of diffusing doublons in a disordered environment. In this model, the interchange between doublon--hole pairs and pairs of opposite-spin singles is controlled by a set of reaction--diffusion (R-D) equations. The R-D model (see Appendix \ref{appendix:rdmodel}) augments the classical continuum diffusion equation for each particle species with a source term that converts a doublon--hole to a single--single combination, only when the local parameters allow this process to be resonant. In particular, our model requires the local energy difference arising from the speckle disorder to lie within a window of width $\sim t$ around $U$ (see Fig.~\ref{fig:Fig1}). The doublon diffusion coefficients are taken to decrease monotonically with disorder strength, as increasingly large local energy differences will inhibit doublon transport throughout the lattice.

The R-D model gives the decay rate $1/\tau$ as a product of the effective diffusion rate $ D_\text{eff}$ and the probability $ P_\text{reaction}$ for a conversion between a doublon--hole pair and two singles ($\uparrow$ and $\downarrow$):
\begin{equation}
 1/\tau = P_\text{reaction} \times D_\text{eff}.
 \end{equation}
\noindent 
The probability of a reaction per site encountered (derived in Appendix \ref{appendix:rdmodel}) is modeled as
\begin{equation}
P_\text{reaction} = p+\exp\left(-U / \Delta\right)\sinh\left(\sqrt{2}z t/\Delta\right), 
\end{equation}
\noindent
where $p$ is a small (i.e., of order 0.01) parameter corresponding to reactions in the clean limit, and $z$ is the lattice coordination number. The effective diffusion coefficient $D_\text{eff}$ gives the rate (linear in time) at which new sites are sampled by any particular doublon. We expect $D_\text{eff}$ to decrease rapidly with disorder since it is controlled by the doublon diffusion constant. Even in the asymptotically localized phase predicted at large disorder (where doublon diffusion vanishes), $D_\text{eff}$ should be supplemented by an additional small velocity that allows sampling of sites within the finite localization length.

The exact dependence of diffusion $D_\text{eff}$ on disorder $\Delta/U$ is not qualitatively important as long as $D_\text{eff}$ decreases monotonically to a very small or vanishing value. In Fig.~\ref{fig:Fig3}, we choose a diffusion coefficient that decreases exponentially with disorder. For this functional form, the best-fit value for $D_{\text{eff}}$ is of order $t/\hbar$ at low disorder (4$\pm$1 $t/\hbar$) and is reduced by a factor of over 100 (to 0.02$\pm$0.008 $t/\hbar$) at the highest $\Delta$. We find that an algebraically decaying diffusion coefficient yields quantitatively similar decay rates at all measured disorder strengths. The combined expression for the decay rate $1/\tau$ within this reaction--diffusion model explains the two relaxation regimes as follows.

In regime 1 ($\Delta \lesssim U$), increasing disorder leads to decreasing decay times. This regime is controlled by the reaction probability $P_\text{reaction}$. In the clean limit, the reaction rate is greatly suppressed by the strong interaction energy $U$, and doublon decay occurs slowly. However, as the disorder strength increases, the tail of the local energy distribution allows for a finite probability of an adjacent site energy difference of order $U$. By compensating the interaction energy $U$ through this mechanism, the disorder produces a quantum resonance that allows the decay reaction to proceed with high probability, thereby resulting in fast doublon relaxation.

In regime 2 ($\Delta \gtrsim U$), increasing disorder leads to increasing decay times. This regime is controlled by the diffusion constant $D_\text{eff}$, which becomes heavily suppressed by localization effects produced by the large disorder. To a lesser extent, the gradual suppression of the reaction probability due to fewer resonances also contributes to the behavior in this regime. Recent theoretical work suggests that, aside from rare region fluctuations, diffusion should vanish at sufficiently large disorder and signal an asymptotic many-body localized phase \cite{Abanin2019}. The diffusion in the 3D interacting system is an unknown function of the disorder ratio to the bandwidth $\Delta/12t$ and interactions $\Delta/U$, but dimensional considerations suggest a reduction when the ratios become of order unity, consistent with the observations Fig. \ref{fig:Fig3}. While the experimental data cannot distinguish between $D_\text{eff}=0$ and $D_\text{eff}$ saturating to a finite small value, we can robustly conclude that diffusion  must be highly inhibited in regime 2.

\section{Conclusion}

Quantum many-body systems involving interactions and disorder can exhibit emergent behavior that, especially in two and three dimensions, is challenging to predict from first principles. By experimentally constructing a DFHM quantum simulator in a 3D optical lattice, we are able to study the dynamical out-of-equilibrium behavior of doublons across vastly different scales of disorder and interactions. Remarkably, we find that constructing a simple model for this extraordinarily complicated system is possible using reaction--diffusion equations that are  analytically solvable.  We devise a numerical approach that captures the same physical effects and shows quantitative agreement with the experiment.

Intriguing open questions remain to be explored. In the intermediate disorder regime ($\Delta \sim U$), the observed fast doublon relaxation may be related to the behavior of ``bad metals'' \cite{Xu2016, Brown2018}, which can be characterized by a lack of conserved excitations \cite{Hartnoll2014}. Because the rapid doublon decay can be understood as a consequence of disorder disrupting the gap in the single-particle spectrum, spectroscopic measurements in this regime could identify a disorder-created pseudogap in the density of states \cite{Lahoud2014, Wang2018}. As a practical tool, the fast disorder-mediated thermalization that we observe also suggests an approach to avoid challenges in the adiabatic preparation of strongly correlated atomic gases \cite{Hung2010}. In the strong disorder regime, our measurements imply a suppression of particle transport as disorder is increased.  However, we are unable to distinguish whether this behavior is a manifestation of slow diffusion or a signature of asymptotic many-body localization. Further experiments could clarify this difference and extend the results of Ref. \cite{Kondov2013a} to the $U/12 t >1$ regime.

\section*{Acknowledgements}
We acknowledge helpful discussions
with Gretchen Campbell and Jason Iaconis. This work is supported by the AFOSR
grant FA9550-18-1-0319; the ARO
single investigator award W911NF-19-1-0210; the NSF
PHY-1820885 and NSF JILA-PFC PHY-1734006 grants; and NIST. It is also supported in part by the AFOSR under grant
number FA9550-17-1-0183 (R.M.N.).

\vspace{1em}
\noindent {\bf Correspondence} and requests for materials
should be addressed to B. DeMarco (email: bdemarco@illinois.edu).

\appendix

\section{Experimental details}
\subsection{Parameters}
\label{appendix:experiment}

\begin{figure}[!hbt]
\centering
\includegraphics[width=0.4\textwidth]{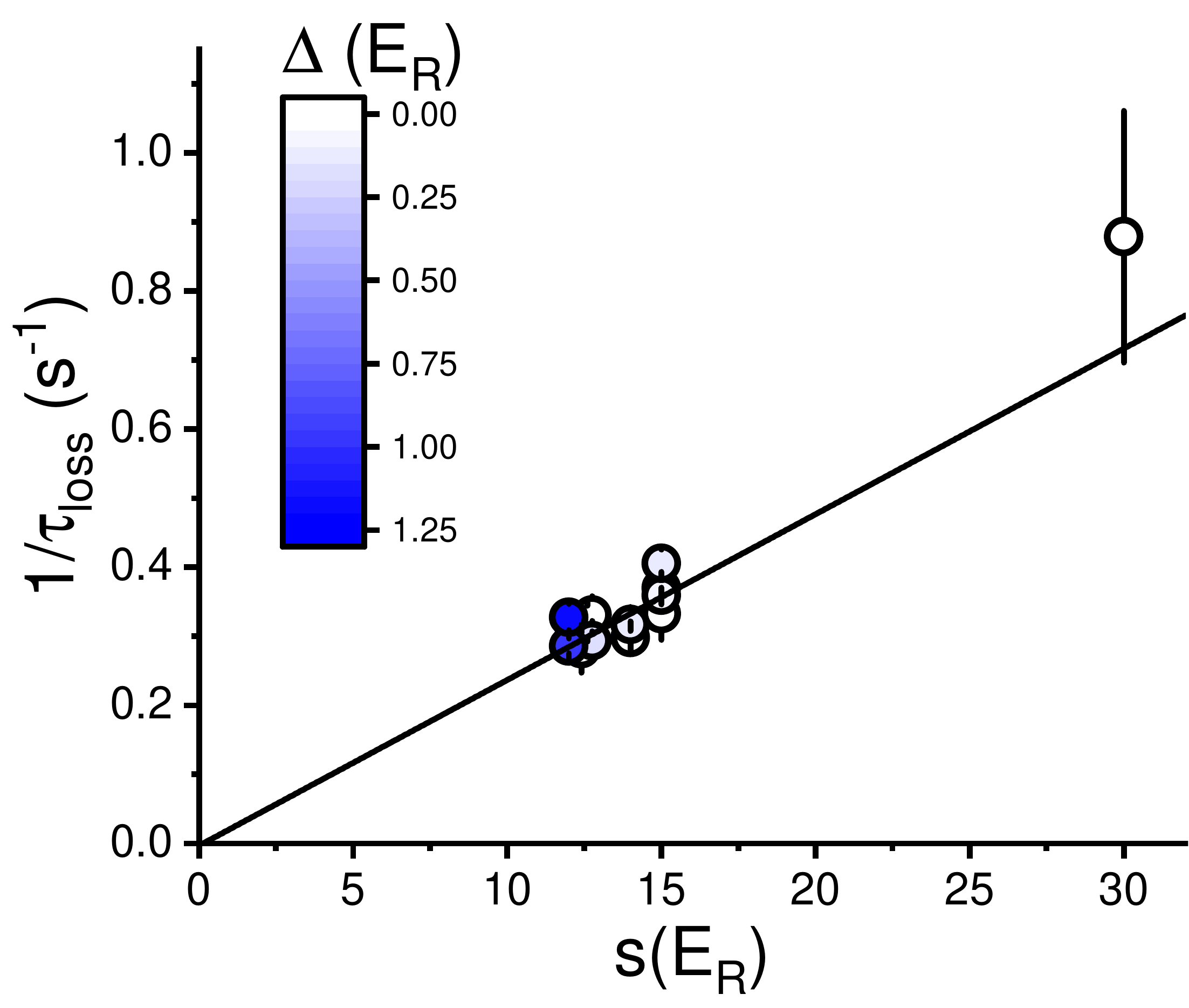}
\caption{Experimental atom number loss rate. The measured loss rate $1/\tau_{loss}$ is shown for various lattice depths $s$ and disorder strengths. Both the lattice depth and disorder are measured in units of the recoil energy of the lattice ($E_R$).  The loss rate scales linearly with lattice depth and is independent of disorder strength. This behavior is consistent  with off-resonant scattering of lattice light as the dominant loss mechanism. The slope of a linear fit (solid line, constrained to have  no intercept) is $0.0238\pm0.0005$ $(\text{s} \cdot E_R)^{-1}$. We use this fit to determine the value of $\tau_{loss}$ used in the fitting model (Eqs. \ref{eq:doublonModel1}-\ref{eq:doublonModel4}) for data sets in which only the doublon population is measured.}
\label{fig:FigS4}
\end{figure}

\begin{figure}
\centering
\includegraphics[width=0.4\textwidth]{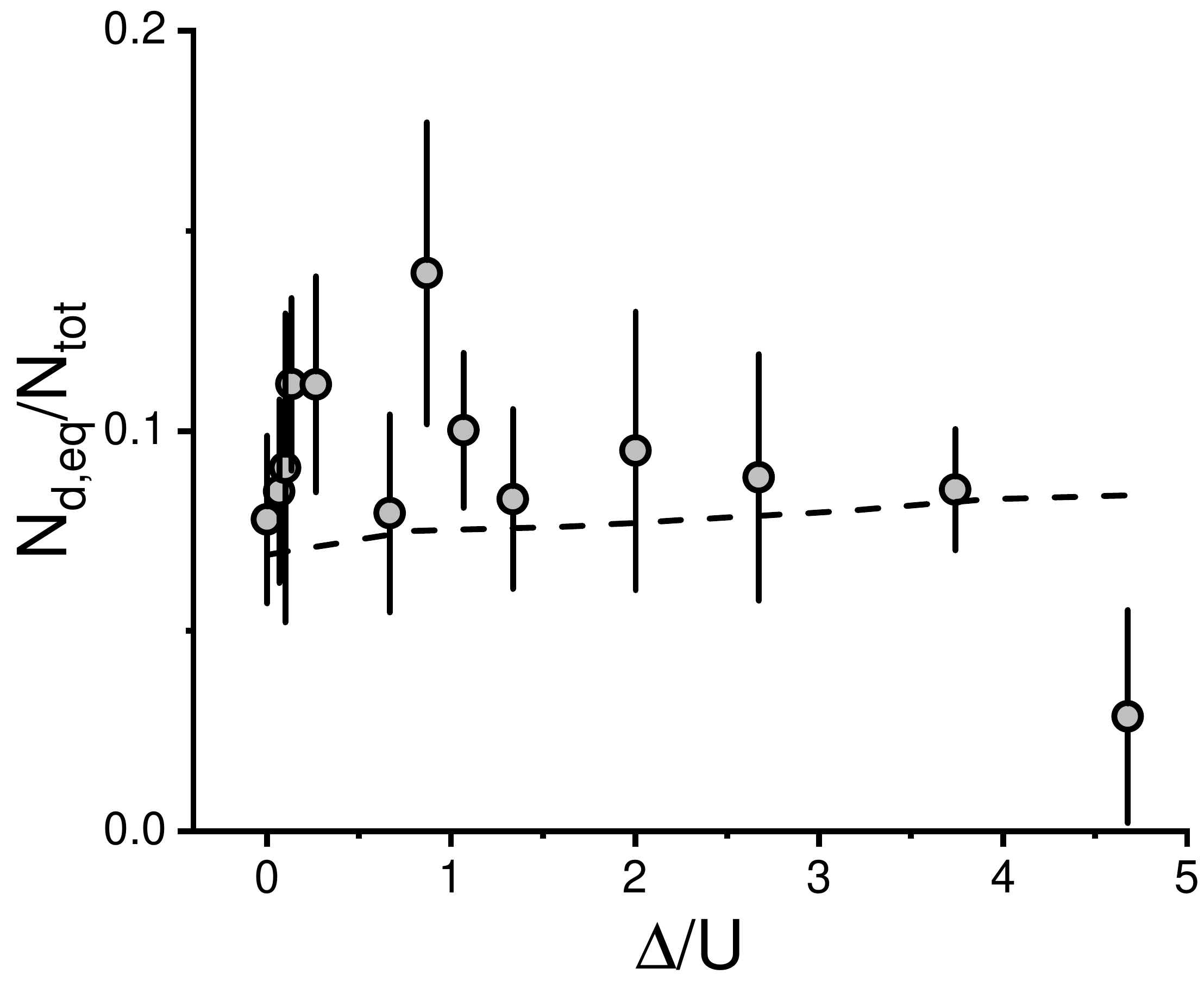}
\caption{Disorder dependence of steady-state doublon fraction ($N_{d,eq}/N_{tot}$), corresponding to the data in Fig. \ref{fig:Fig3}. The steady-state doublon population $N_{d,eq}$ is extracted from the fits to the doublon data, and corrected for finite imaging efficiency, while the total number in one spin $N_{tot}$ is measured in separate experimental runs. Error bars are the standard error of the fit. The dashed line is an atomic limit calculation for equilibrium at the initial temperature and number of atoms.}
\label{fig:FigS5a}
\end{figure}
We use a balanced mixture of the $\left|F=9/2, m_F=-9/2\right\rangle$ and $\left|9/2, -7/2\right\rangle$ $^{40}$K spin states, which are denoted as $\left|\uparrow \right\rangle$ and $\left|\downarrow \right\rangle$. Typical starting conditions are 100,000 atoms at $T/T_F=0.4$ ($T=$ 100 nK). The atoms are initially confined in a 1064 nm optical trap with the interactions tuned to be attractive (scattering length $a_s$ = $-75.8$ $a_0$, where $a_0$ is the Bohr radius) using the Feshbach resonance near 202.1~G \cite{Loftus2002}. We load the atoms into a 3D cubic optical lattice with lattice depth $s=12$~$E_R$, where $E_R=h^2/(2m\lambda^2)$ is the recoil energy ($m$ is the atomic mass and $\lambda=782.2$~nm is the lattice wavelength). The state of the gas before the interaction quench is fixed for the data in this study. The quench is realized by increasing the magnetic field by 7~G over $100$ $\mu$s, after which $a_s$ = $61.7$~$a_0$ and $U/12t=1.8$ (see Appendix \ref{appendix:interaction} for data at other interaction strengths).

Immediately after the quench, the disorder is turned on over $1$~ms  with a strength ranging from $0$--$1.2$~$E_R$. The disorder is generated by a 532~nm speckle laser beam created using a holographic diffuser \cite{Kondov2013a, Kondov2011}. The strength of the disorder $\Delta$ is tuned by controlling the 532~nm optical power and is equal to the standard deviation of the disorder potential. The disorder ramp rate is chosen to be adiabatic relative to the lattice band gap, but fast compared to the doublon decay rate in the absence of disorder. The rapid turn-on of the disorder may induce disorder-dependent heating of the atoms. Using an indirect thermometry method, in which we measure the size of the gas after adiabatic turnoff of the lattice and disorder potential, we infer that this effect changes the temperature by less than 20\%.

After a variable hold time $t_{hold}$, the lattice potential depth is quickly ramped to 30~$E_R$ to freeze the density distribution, and the disorder is removed by turning off the 532~nm light. The doublon population is measured by ramping the magnetic field across the Feshbach resonance to associate doublons into molecules \cite{Jordens2008}, selectively transferring the $\left|\downarrow \right\rangle$ atom in each molecule to the ancillary $\left|9/2, -5/2\right\rangle$ state with an rf pulse, and imaging only the atoms in the ancillary state. Before imaging, we remove all atoms in the $\left|\downarrow\right\rangle$ and $\left|\uparrow \right\rangle$ spin states using a combination of rf sweeps and resonant light pulses. Then,  the atoms in the ancillary state are transferred to the $|\uparrow\rangle$ state to be imaged on a closed transition. We can also image only the $\left|\downarrow \right\rangle$ atom singles instead of doublons by changing the rf pulse to be resonant with the spin transition for free atoms instead of molecules.

\subsection{Fitting model}

We extract $\tau$ using a fit to a rate model:
\begin{gather}
\dot{N}_{d}=\dot{N}_{d, neq}+\dot{N}_{ d, eq}, \label{eq:doublonModel1}\\
\dot{N}_{d, neq}=-\frac{1}{\tau}N_{d, neq}-\frac{1}{\tau_{loss}}N_{d, neq}, \label{eq:doublonModel2}\\
\dot{N}_{d, eq}=-\frac{1}{\tau_{loss}}N_{d, eq}, \label{eq:doublonModel3}\\
\dot{N}_{s\downarrow}=-\frac{1}{\tau_{loss}}N_{s\downarrow}+\frac{1}{\tau}N_{d, neq}.
\label{eq:doublonModel4}
\end{gather}
\noindent
This  model describes a non-equilibrium population of doublons $N_{d, neq}$ that dissociate to create $N_{s\downarrow}$ singles $\left| \downarrow \right \rangle$ (along with undetected $\left| \uparrow \right \rangle$ atoms) at a rate $1/\tau$, a steady-state population of doublons $N_{d, eq}$, and overall number loss at rate $1/\tau_{loss}$. For the longest doublon lifetimes sampled in this work, we perform a simultaneous fit to the changes in the doublon and singles populations to extract $\tau$, while for the shorter lifetimes we find that the singles provide no additional constraint. Measurements of $\tau_{loss}$, which primarily depends on the lattice depth, are shown in Fig. \ref{fig:FigS4}.

The fit values for $N_{d,eq}$ are shown in Fig. \ref{fig:FigS5a}. Unlike the measurements of $\tau$ presented in the main text, these are only physically meaningful given knowledge of the imaging efficiency. Experimentally, we find that this efficiency is lower (by approximately a factor of 2) for measurement of doublons than for singles, which we attribute primarily to inelastic decay of Feshbach molecules during the imaging procedure. Therefore, to arrive at an equilibrium doublon fraction, we scale the measured number by the initial total atom number (in one spin state) and correct for the unequal imaging efficiencies of doublons and singles by comparing their respective population changes during decay. The resulting steady-state doublon fraction is comparable to the result from an atomic limit calculation at equilibrium (dashed line) \cite{DeLeo2011}. This calculation neglects the entropy generated by the quench, which is expected to be substantially smaller than the entropy present initially from finite temperature.

\subsection{Dependence of $\tau$ on $U/12t$}
\label{appendix:interaction}

\begin{figure*}[!thb]
\centering
\includegraphics[width= \textwidth]{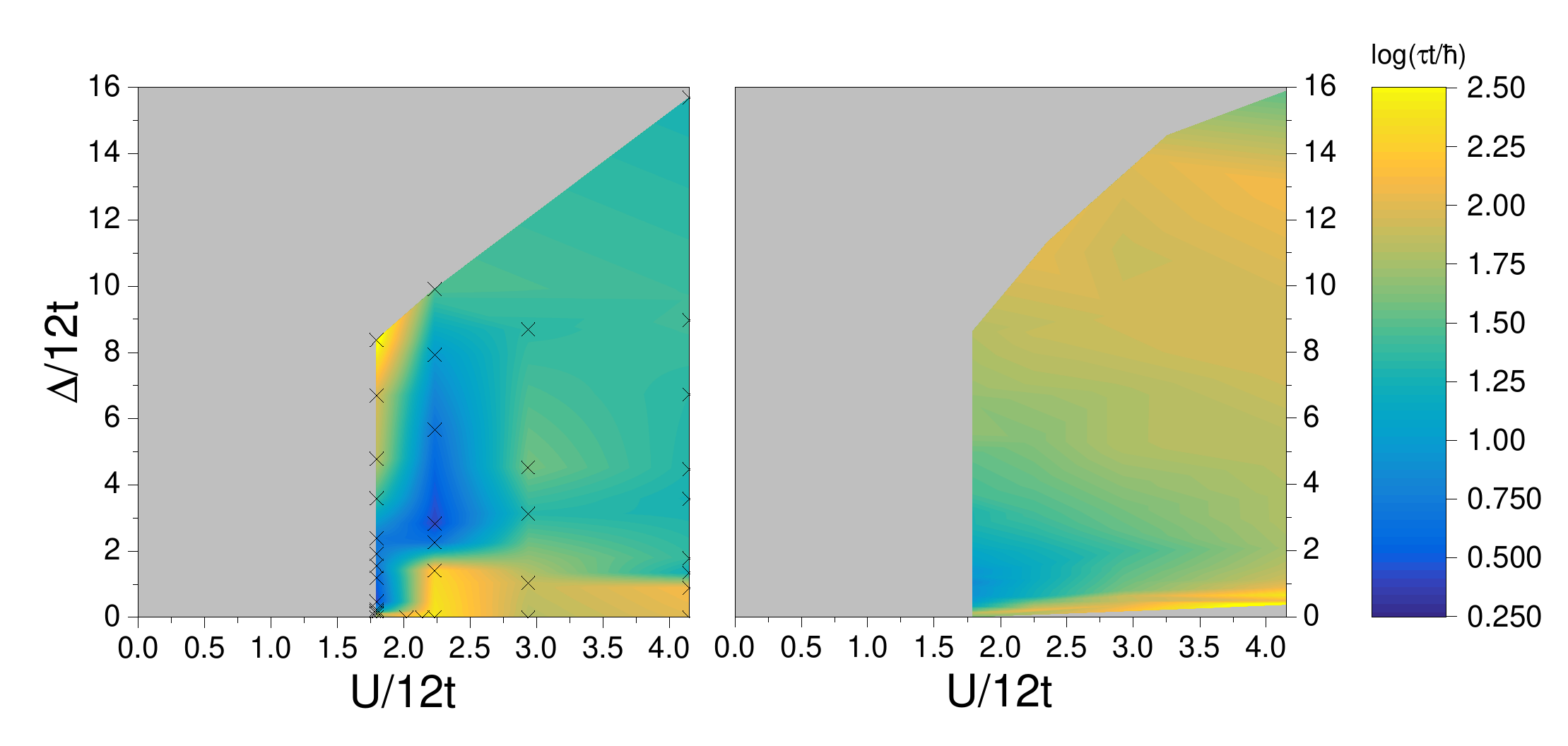}
\caption{Dependence of measured $\tau$ on $U/12t$. Left: experimental data. The lattice depth is tuned from 12--15~$E_R$ to vary $U/12t$ by a factor of approximately 2.5. The crosses represent experimental points, which are interpolated to create the colormap. Right: GDTWA numerical results.}
\label{fig:EDInteract}
\end{figure*}

By changing the lattice depth, we measure the dependence of the relaxation on interaction strength, allowing us to map out an experimental relaxation phase diagram. We vary $s$ from 12--15 $E_R$, deep into the doped Mott insulator regime at equilibrium \cite{Semmler2010}. For lower lattice depths, the effect of the quench becomes too small to accurately extract $\tau$. The resulting data, along with the corresponding numerical GDTWA calculations, are shown in Fig. \ref{fig:EDInteract} (we omit the reaction-diffusion model, as it provides a model and fit, rather than an independent prediction).

For all values of $U$ we observe the initial disorder-driven decrease in relaxation time, both in experiment and numerics. In both experiment and numerics, we also see a strong dependence on disorder at low $U$, while at high $U$ the differences become washed out. However, differences between experiment and numerics are also evident. For high $U$ the regime of increasing relaxation time appears to recede beyond accessible values of $\Delta$ in experiment.

For the strongest interactions ($U/12t=2.9,4.2$), the measured lifetime in the clean limit is likely limited by technical imperfections; the predicted elastic doublon decay lifetimes for these points (using the scaling of Ref. \cite{Strohmaier2010}, which is consistent with our measurements at lower $U/12t$) are one and nine seconds, respectively. However, the numerics also do not capture this limit, in which the decay mechanism is via higher-order processes \cite{Sensarma2010}. More generally, the experimental data may exhibit many-body effects due to the finite filling and also finite temperature effects, which are interesting subjects for future work.

\section{Generalized Discrete Truncated Wigner Approximation}
\label{appendix:gdtwa}

\begin{figure*}[!htb]
\centering
\includegraphics[width=0.9\textwidth]{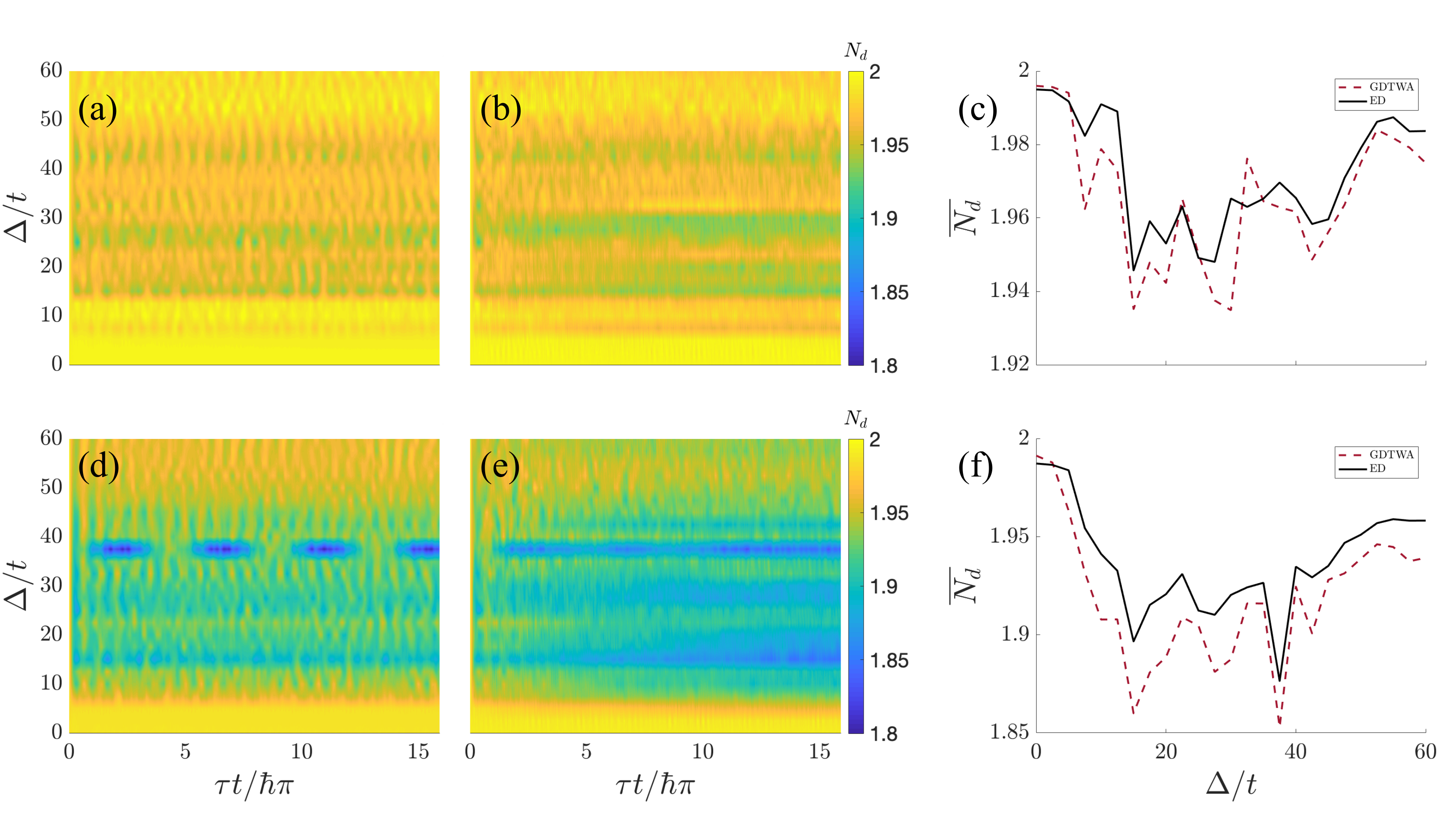}
\caption{GDTWA benchmarking in 1D. A comparison of the total doublon number dynamics obtained from exact diagonalization of the DFHM(a,d) and GDTWA (b,e) is shown for a periodic 1D chain of eight lattice sites. These simulations were carried out with $U/12t = 2.5$, and the results were averaged over 100 disorder realizations for each disorder strength. The top row (a-c) corresponds to an initial loadout of two doublons and no singles, while the bottom row (d-f) corresponds to an initial loadout of two doublons and one single of each flavor. Panels (c) and (f) give the doublon population time-averaged over the last half of the time interval shown. For system sizes accessible by exact simulation, the initial doublon population quickly jumps to the steady state, without any further relaxation features observable on longer timescales.}
\label{fig:FigS6}
\end{figure*}

\begin{figure}
\centering
\includegraphics[width=0.45\textwidth]{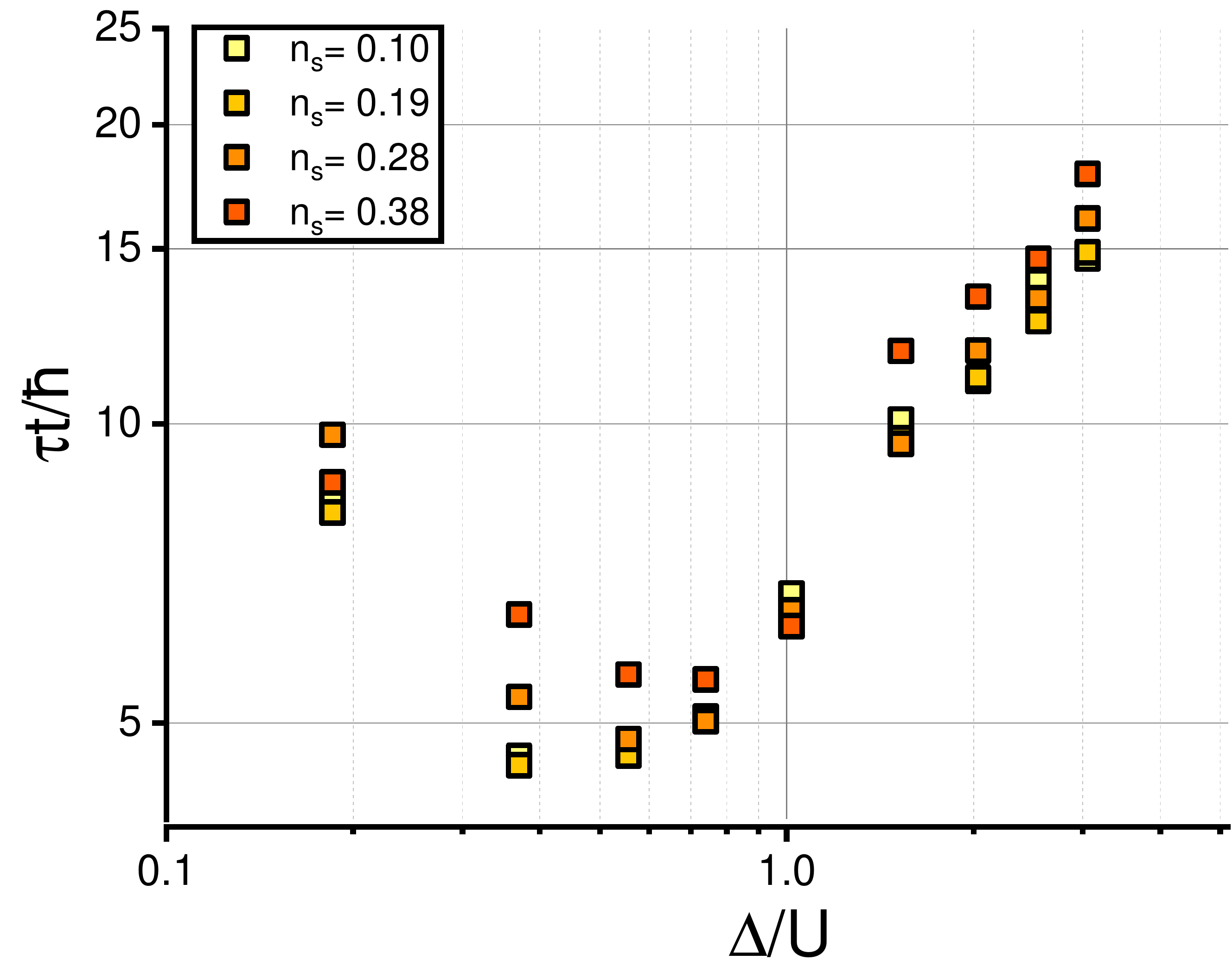}
\caption{GDTWA variation of decay times with singles density. A comparison of GDTWA decay times  at selected values of $\Delta/U$ for a $6\times 6 \times 6$ lattice with different initial singles densities. These results were averaged over 50 disorder/initial configuration realizations, with each configuration also averaged over 50 GDTWA trajectories. Here, $U/12t = 1.8$, and the initial doublon density remains fixed at 11\%. While the steady-state population depends strongly on the initial singles density, the robustness of the decay time to this parameter allows comparison with experiment, despite the spatially varying density in the trap.}
\label{fig:FigS7}
\end{figure}

\begin{figure}
\centering
\includegraphics[width=0.45\textwidth]{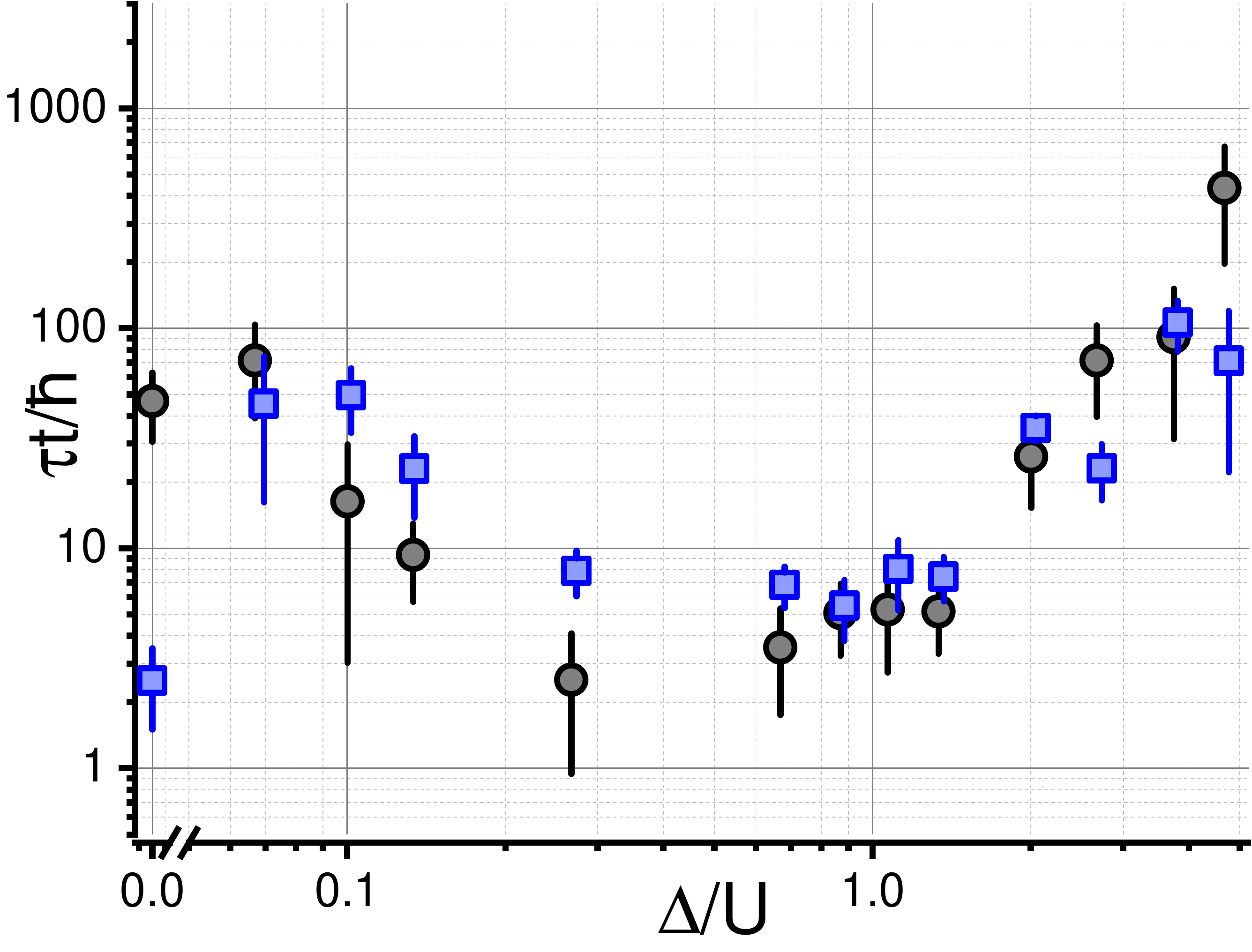}
\caption{GDTWA relaxation times from biased fits. A comparison between experimental (black circles) and GDTWA decay times (blue squares) is shown.  The GDTWA decay times are extracted utilizing a (disorder-dependent) sampling function consisting of delta peaks at the experimental hold times, weighted by the appropriate experimental uncertainty. The error bars show the fit standard error, obtained from a linearized curvature analysis.}
\label{fig:FigS8}
\end{figure}

To numerically simulate doublon dynamics, we adapt the generalized discrete truncated Wigner approximation (GDTWA) to our system \cite{Zhu2019}. We begin by choosing a tensor product structure for the Hilbert space; for the DFHM, we let each subspace correspond to the four-dimensional Hilbert space for each lattice site with basis $\left\{\ket{\uparrow\downarrow},\ket{\uparrow},\ket{\downarrow},\ket{0}\right\}$. We then invoke a factorization of the density matrix as a function of time, $\hat{\rho}(\mathfrak{t})$, along this tensor product structure, resulting in
\begin{align}
    \hat{\rho}(\mathfrak{t}) = \bigotimes_j \hat{\rho}_j(\mathfrak{t}).
\end{align}
\noindent
This ansatz allows us to retain information regarding the strong on-site correlations between spin species responsible for doublon formation, while treating tunneling induced correlations between sites in a semiclassical manner. However, this method also implicitly invokes a replacement of the anti-commuting fermionic operators with commuting hardcore boson operators. Nonetheless, we expect this to be a good approximation, given that the doublon population relaxation dynamics are expected to be  highly insensitive to quantum statistics in the  dilute regime explored in the experiment.

In principle, it is possible to increase the size of each Hilbert space in our product state factorization ansatz (e.g., by including multiple lattice sites within each density matrix factor) and thus presumably increase the accuracy of the simulated dynamics. However, the computational cost associated with randomly sampling over disorder distributions precludes the use of this strategy in this work. Already, a cluster of only two sites requires a description via 256 phase space variables per cluster. We have performed limited simulations for select parameters using this clustering method, and generally found that the resulting decay times are robust, though the equilibrium doublon values are altered slightly.

The factorization approximation directly results in a set of mean-field-like equations for variables $\lambda_j^{\alpha}(\mathfrak{t}) \equiv \langle\hat{\Lambda}_j^{\alpha}(\mathfrak{t})\rangle$, where the 
Hermitian operators $\hat{\Lambda}_j^{\alpha}$ correspond to the set of generalized Gell-Mann matrices (GGM) for SU(4) (along with the identity matrix) with index $\alpha$ for each lattice site $j$, which form  a complete orthonormal basis under a Hilbert-Schmidt norm for the observables on $j$. The  resulting mean-field equations are given by:
\begin{equation}
    \frac{d\lambda_j^{\alpha}}{d \mathfrak{t}} = i\left[\sum_{\beta}M_j^{\alpha\beta}\lambda_j^{\beta}(\mathfrak{t}) + \sum_{\beta,\beta',j'} C_{jj'}^{\alpha\beta\beta'}\lambda_j^{\beta}(\mathfrak{t})\lambda_{j'}^{\beta'}(\mathfrak{t})\right]
\end{equation}
where, for a generic Hamiltonian written as  $\hat{H} = \sum_j \hat{H}_j + \sum_{jj'}\hat{H}_{jj'}$, we have $M^{\alpha\beta}_j = \mathrm{Tr}\left[\hat{\Lambda}^{\beta}_j [\hat{H}_j,\hat{\Lambda}^{\alpha}_j]\right]$ and $C^{\alpha\beta\beta'}_{jj'} = \mathrm{Tr}\left[\hat{\Lambda}^{\beta}_j\hat{\Lambda}^{\beta'}_{j'} [\hat{H}_{jj'},\hat{\Lambda}^{\alpha}_j]\right]$. We use the equations arising from this approximation (which lead to trivial mean-field dynamics) to dynamically evolve multiple trajectories randomly sampled from a phase space \cite{Zhu2019}, with a sampling distribution determined by the initial product state of the system. Specifically, if each observable basis element has eigendecomposition $\hat{\Lambda}^{\alpha}_j = \sum_a \lambda^{\alpha,[a]}_j\hat{P}^{\alpha,[a]}_j$ for eigenvalue $\lambda^{\alpha,[a]}$ and corresponding eigenspace projector $\hat{P}^{\alpha,[a]}_j$, then for each trajectory we set $\lambda^{\alpha}_j(0) = \lambda^{\alpha,[a]}_j$ with probability $p^{\alpha}_j(a) = \mathrm{Tr}\left[\hat{\rho}_j(0)\hat{P}^{\alpha,[a]}_j\right]$.

In our simulations, we neglect the relatively small disorder in the tunnelling and interaction parameters. We also neglect the presence of the harmonic confining trap: assuming diffusive motion, the doublons in the trap center will not travel an appreciable distance towards resonant regions in the trap edges before decaying. We expect both of these features to result in small shifts to the effective disorder strength and diffusion coefficients, but not to affect the main features of our results.

As we are interested in only the total doublon population, we let each sampled trajectory correspond to a different quenched disorder realization and a different random initial product state configuration respecting the lattice filling (see comparisons with exact results in Fig.~\ref{fig:FigS6}). Each dynamical curve in Fig.~\ref{fig:Fig2} and the dynamical curves corresponding to the decay times in Fig.~\ref{fig:Fig3} results from averaging 500-1,000 such trajectories in a $4\times4\times4$ lattice with periodic boundary conditions, with a randomly distributed initial loadout of fixed doublon and singles densities (with an even mix of singles spin orientation). Additionally, we multiply the simulation results by a phenomenological particle loss factor using the measured loss rate when comparing to the experiment (see Fig.~\ref{fig:Fig2}). In benchmarking, we have generally found that the resulting decay times are robust to variations in the singles density at all disorder strengths (see Fig.~\ref{fig:FigS7}). The insensitivity to density enables comparison with the experimental doublon decay timescales, despite the variation in singles density across the trap. For extracting decay times in Fig.~\ref{fig:Fig3}, we utilize the same initial doublon and singles densities (for all disorder strenghts), which are representative of the typical initial densities measured in the trap center (we set the initial doublon density to $0.11$, and the total particle density to $\langle n \rangle = 0.59$).

While the numerical simulations at moderate and late times are consistent with a single exponential decay, an initial rapid dephasing process at very short times deviates from this prediction. Such a feature is not resolvable in the experimental data, which have an approximately thermal initial state rather than a product state, and in which very short time dynamics are not accessible due to the finite quench speed. In order to extract consistent decay times, we perform two different fitting methods on the simulation data. The first method is optimized to accurately capture the initial decay of the numerical data, while the second is optimized for comparison to the experiment. In the first method (displayed in Fig.~\ref{fig:Fig3}), we fit the dynamical curves to an exponential decay with a nonzero steady-state, and we constrain this function to start at the exact initial doublon number. This technique leads to decay times that incorporate the effect of these rapid early time decay dynamics (see Fig.~\ref{fig:Fig2}). An unconstrained fit virtually ignores the early time regime and is more heavily influenced by the long period of slow, exponential relaxation at late times, leading to an underestimate of the decay time and a significant underestimate of the initial doublon number. In the second method, which is displayed in Fig.~\ref{fig:FigS8}, we only use simulation data at times corresponding to the experimental measurements. This latter method takes into account experimental bias from the selected hold times, thus providing a consistency check between simulation and experiment. This technique produces generally better quantitative agreement in the crossover regime.

In contrast to the decay time, the steady-state doublon fraction is strongly dependent on the singles density and generally increases as more initial singles are added to the system. To compare with the dynamical data from the experiment in Fig.~\ref{fig:Fig2}, for each disorder strength we perform simulations for various singles densities (each differing by $\approx 0.03$) and select the simulation corresponding to the best weighted least-squares fit to the experiment. Keeping the initial doublon density fixed at $0.09$, we find best-fit total particle densities $\langle n \rangle = 0.22$, $0.53$, $0.53$, and $0.41$ for Fig. 2(i), (ii), (iii), and (iv), respectively. While this best-fit singles density is generally representative of the initial densities measured in the experiment over a wide range of disorders, for small disorders the best-fit singles density are well below reasonable values for the singles density in the experiment, to the point where a vanishingly small initial singles density cannot accommodate the measured change in doublon density. This is the result of various decay processes in the clean system not being captured by GDTWA, resulting in a higher equilibrium doublon fraction. However, we still find that the extracted decay timescales are generally consistent with experiment in this regime.

\section{Reaction-Diffusion Model}
\label{appendix:rdmodel}
In this model, we let each particle species obey a R-D equation, e.g.\ for the doublon number $n_d(\mathfrak{r},\mathfrak{t})$, as a function of space $\mathfrak{r}$ and time $\mathfrak{t}$,

\begin{equation}\partial_{\mathfrak{t}} n_d = D_d \nabla^2 n_d + S(\mathfrak{r}) (n_{s\uparrow}n_{s\downarrow}-n_d n_h) . \end{equation}
\noindent
The source term $S(\mathfrak{r})$ has contributions from quantum fluctuations and disorder. Here we focus on the latter contribution, relevant when disorder is not too small. $S(\mathfrak{r})$ is modeled by a small constant clean-system reaction probability $p$, together with a Boolean step function signifying resonance with local energy difference $\mu(\mathfrak{r})$,

\begin{equation} S(\mathfrak{r}) = p + \Theta\left(\gamma - |\mu(\mathfrak{r})-U|\right),\end{equation}
\noindent
where $\gamma$ is the width of the resonance and $\mu(\mathfrak{r})$ is the difference of two independent random local energies from the speckle distribution,

\begin{equation}{\wp}(\mu) = \frac{1}{2\Delta} \exp{\left(-|\mu|/\Delta\right)}. \end{equation}

The resulting $S(\mathfrak{r})$ distribution yields a reaction ($S\approx 1$) with probability

\begin{align}P(\{S=1\}) &= p + \int_{U-\gamma}^{U+\gamma}{\wp}(\mu) d\mu \nonumber \\ &= p + \exp(-U/\Delta) \sinh(\gamma/\Delta), \end{align}
\noindent
which is also the average of $S(\mathfrak{r})$. We expect $\gamma \approx \sqrt{2}z t$ for lattice coordination number $z$ ($z = 6$ for a cubic lattice in 3D), though the exact form is unimportant for the general conclusions we wish to draw. We also find that the form of $S(\mathfrak{r})$ is qualitatively similar to the form obtained by convolving the speckle distribution with a heavy-tailed Lorentzian resonance peak.

To proceed, let us work to linear order in $n_d$ (controlled in the dilute limit) and take $t \ll \Delta$, arriving (with $D=D_d$) at a linear differential equation with an inhomogenous random source term,
\begin{equation}\partial_{\mathfrak{t}} n_d(\mathfrak{r},\mathfrak{t})= D \nabla^2 n_d(\mathfrak{r},\mathfrak{t}) - S(\mathfrak{r}) n_d(\mathfrak{r},\mathfrak{t}). \end{equation}
\noindent
The homogeneous solution to the heat equation of a point particle in $d = z/2$ dimensions is $n(\mathfrak{r},\mathfrak{t}) = (4\pi D \mathfrak{t})^{-d/2} \exp{\left(-\mathfrak{r}^2/4D \mathfrak{t} \right)}$. 
The linear diffusion range grows as $\sqrt{2dD \mathfrak{t}}$, which is also proportional to the number of sites sampled when $d=1$. In 3D, however, the number of sites sampled grows linearly in time as $N_S =  \alpha D \mathfrak{t}$ with $\alpha \approx 1.32$. This yields a decay constant

\begin{equation} \Gamma = \alpha D(\Delta) \left[p + \exp(-U / \Delta) \sinh\left(\sqrt{2}zt/\Delta\right)\right]. \end{equation}
\noindent
For a diffusion constant of the form $D(\Delta) = D_0 \exp{(-\Delta/\Delta_0)}$, we fit the model to the experimental data with unknowns $p$, $D_0$, and $\Delta_0$, resulting in the line shown in Fig. \ref{fig:Fig3}. The best fit parameters are $p = 0.011\pm 0.007$, $\hbar D_0/t = 4 \pm 1$, and $\Delta_0/U = 0.88 \pm 0.09$.

\bibliography{doublonlibrary_v2}

\end{document}